\documentclass[aps,prb,floats,twocolumn]{revtex4}
\usepackage{epsfig}

\begin{document}
\title{Long range superexchange - an exchange interaction through empty bands} 
\author{S. Schwieger and W. Nolting}
\begin{abstract}
We derive a generalization of the RKKY interaction to semiconductors
using perturbation theory on a non-degenerated two-impurity Anderson model. In metals the interaction is mediated by
excitations of free carriers over the Fermi-energy. In semiconductors, where no
carriers are present, the only possible excitations are those of the localized
impurity electrons (or holes) themselves. Thus a possible interaction is
closely related to superexchange. We find an oscillating
anti-ferromagnetic spin-spin coupling due to impurity electron (hole)
excitations. By treating the coupling through empty bands
(superexchange) along the same route as carrier mediated interactions (RKKY)
it is easy to compare these two kinds of spin-spin coupling. The interaction derived here is of special interest for diluted
magnetic semiconductors.

\end{abstract}
\maketitle
Usually superexchange is formulated within a cluster model consisting of
three sites: two cation orbitals are partly filled thus forming an
effective spin moment and one intermediate anion orbital is completely
filled. In fourth order perturbation theory the resulting spin-spin
interaction between the cation sites reads ($180^\circ$ Mn-O-Mn):
\begin{equation}
J=-\frac{2V^4}{\Delta T_0^2}(U^{-1}+\Delta T_0^{-1}),
\label{1}
\end{equation}
where $\Delta T_0$ is the difference between the ground state and a
configuration where one electron is transferred from the anion to the
cation. $U$ is the on-site Coulomb interaction at the cation and $V$ is
 the hybridization between both kinds of electrons.\\
However it is out of question, that the superexchange has also a long
range component. The latter is very important for diluted magnetic
semiconductors, especially for doped $(II,Mn)VI$ semiconductors. A
competition between RKKY interaction and superexchange is typical
for these materials \cite{KLM01,KLM00,FCW01,KAC01}. Nearest neighbor superexchange leads to local spin singlets
that reduce the effective concentration of Mn spins. The superexchange between more
distant pairs of Mn gives an anti-ferromagnetic coupling that competes
with RKKY as soon as free carriers are present, which may be generated by
doping with N in $(II,Mn)VI$ semiconductors or are present from the very
beginning in $(III,Mn)V$ semiconductors. To get a qualitative picture of
the interplay between the different exchange interactions it is
convenient to have some simple limiting expressions of the mechanisms at
hand. These expressions should give an idea of the dependence of the
exchange mechanisms on certain model parameters like e.g. the
intraatomic exchange coupling or the local Coulomb repulsion of Mn
d-electrons and the (sp)-d hybridization. For superexchange Eq. (\ref{1}) shows
these dependencies.\\ Concerning the long range part of the
interaction additional informations are crucial, i.e. the dependence on
the inter spin distance $\Delta$ and the dependence on the electronic band
structure of the host material. Both properties can not be deduced from
Eq. (\ref{1}) or from expressions derived from any other cluster
model. It is the intention of this paper to derive an expression for the
long range component of superexchange, which is well comparable to the RKKY
interaction. To this aim we will study a toy model and adjust the
parameters in such a way, that the so to say "standard RKKY situation"
is recovered. That means two isolated spins should be located at a
certain distance in a host material that is described by a
nondegenerate uncorrelated band. For this limiting case we will then
apply fourth order degenerated perturbation theory.\\
The paper proceeds in the following way: First the most important
indirect exchange mechanisms are briefly reviewed and discussed. We will
concentrate on superexchange, RKKY, Bloembergen-Rowland interaction and a mechanism
similar to the latter as well as to superexchange. Compact expressions for the last
three coupling mechanisms are discussed. Then we will introduce the toy
model and adjust the parameters in such a way, that we reach the best
comparability to the RKKY expression. In the next step we will derive an
expression for superexchange that will be exact in fourth order perturbation theory
for the prepared model situation. Since the toy model establishes a well
defined limit for more complex calculations, this expression can be used
as a check for certain approximations. For demonstration we will compare
a work of Larson et.al. \cite{LHE88} with our result. Furthermore the result should
give a vivid idea of the distance and band structure dependence of
superexchange. To this purpose we evaluate the superexchange expression numerically for some
simple model lattices.
\section{Indirect exchange mechanism}
Indirect exchange mechanisms, i.e. effective spin-spin couplings usually
between local spins at cation sites mediated by diamagnetic anions where
intensively discussed in the 50th. e.g. by Anderson \cite{AND59}, Goodenough and
others\cite{GOO58}. The main goal of these studies was to understand magnetism in
insulators such as  MnO and to justify the use of the Heisenberg model
for this class of materials. The works where primarily concerned with
the leading interaction of spins in adjacent lattice cells and
consequently a lot of cluster models where adopted.\\
A different topic is the effective spin-spin interaction in metals. Here
the interaction is mediated by free carriers. In the language of
perturbation theory these carriers are virtually excited over the Fermi
energy, which results in a spin-spin coupling that oscillates in sign
and altitude in dependence of the inter spin distance. Such an
interaction is usually called RKKY coupling. It was first proposed by
Ruderman and Kittel for nuclear spins \cite{RUK54} and later generalized to electronic
spins. It is often discussed e.g. in heavy fermion systems. In contrast
to indirect spin exchange in insulators RKKY interaction is formulated
within a band picture and it's dependence on the spin-spin distance is
well known.\\
In semiconductors both mechanisms, i.e. virtual excitations of carriers
and non-carriers, may be important and even compete with each
other. However for a lot of materials the restriction to spins of
neighboring lattice cells, typical for insulators, is not a good
approximation any more. Let us mention Europium chalcogenides, where at
least the next nearest neighbor cell is important or diluted magnetic
semiconductors (DMS). Therefore a band formulation that is analogous to
RKKY is also desired for interactions caused by excitations of
non-carriers, like e.g.~superexchange.\\
Such interaction types are widely discussed especially for
semiconductors. Let us start with Bloembergen Rowland\cite{BLR55} interaction, that
is the band analogue to the process described for clusters in
Ref. \cite{AND59} in Eq.. (29;30). A valence electron is
virtually excited at site 1, both the electron and the hole are
transferred and recombine at site 2. The spin of the electron and the
hole are coupled to local spins at site 1 and 2 by an intraatomic
inter-orbital potential. This interaction is believed to be responsible
for the magnetic interactions in Eu-chalcogenides \cite{LIU80} and also
discussed for DMS \cite{BAL79}. A similar interaction (Eq. 23 of
Ref. \cite{AND59} is discussed by Litvinov and Dugaev for
(III,Mn)V DMS \cite{LID01}. In the following we will call the this interaction
"impurity induced Bloembergen-Rowland interaction" since the impurity
electrons (Mn-d electrons in (III,Mn)V DMS) are virtually excited
instead of valence electrons.\\
For all couplings discussed so far there is a "standard expression"
usually derived in perturbation theory for some toy model that describes
pure basic conditions for the respective interactions.\\
For RKKY this set up consists of two spins which are locally coupled to
an uncorrelated partly filled electron band by an intraatomic spin-spin
interaction $J_{pd}$. The same holds for impurity induced Bloembergen-Rowland
interaction, just that the "spin" is now described by partly filled
localized electron orbitals and the electron band is empty instead of
partly filled. The incomplete filling may be due to a strong on-site
Coulomb repulsion. For the classical Bloembergen-Rowland interaction we need again two
spins and a completely filled valence- as well as an empty conduction
band. Again, the spin of the electrons are coupled to the local spins by
an intraatomic interaction.\\
The resulting expressions for RKKY and Bloembergen-Rowland interaction
in perturbation theory read (natural units):\\
RKKY:
\begin{equation}
J(\Delta)=-\frac{J_{pd}^2}{2N^2}\sum_{k,k^\prime}\frac{\cos((k^\prime-k)\Delta)}{\epsilon_{k^\prime}-\epsilon_k}
\end{equation}
The sum runs over all $k$ within the Fermi sphere and all $k^\prime$
that are located outside. 
For parabolic bands one finds for the Bloembergen Rowland interaction:
\begin{equation}
J(\Delta)=-\frac{J_{pd}^2m^2 \Delta T_0}{\pi^3 \Delta^2}K_2(2r/r_0)\quad
r_0=(2m\Delta T_0)^{-1/2}
\end{equation}
with the band gap $\Delta T_0$, the inter spin distance $\Delta$ and the
reduced effective electron mass $m$. $K_2$ is the Mac-Donald function.
($K_2\sim 1/\Delta^2 \,for\,\, r\ll r_0,\, K_2~\sim~
\Delta^{-3/2}~e^{2\Delta/r_0} \,for\, r\gg r_0$).\\
The same holds for the impurity induced Bloembergen-Rowland interaction,
where $\Delta T_0$ is now the energy difference between the impurity
and the conduction band and m is now the effective electron mass in the
conduction band\cite{LID01}. In the next section we want to treat superexchange
and derive a similar expression for this interaction.\\
\section{Model description for superexchange}
The model that can describe a "pure" version of the long range superexchange
should be similar to the above mentioned models, especially to the model
for RKKY interaction. Thus the competition between the latter and superexchange can
be studied. Our model consists of two impurity sites with an effective
spin moment. This moment is due to partly filled localized orbitals,
which is realized by a strong on-site Coulomb repulsion $U$ at the
impurity orbitals. Furthermore there is an free electron band ,
described by the dispersion $\epsilon_k$, that is energetically
separated from the impurities by an energy $\Delta T_0$. The chemical
potential is located between the impurity orbital and the band. The
latter is thus completely empty in the unperturbed ground state. The
impurity orbitals and the band are kinetically coupled by a local
hybridization $V$. The latter will constitute the perturbation in the
following calculation. This is a minimal set to study superexchange. Therefore other
features, like e.g. intra-orbital Coulomb exchange, are not taken into
account.\\\\
The Hamiltonian for the described model reads:
\begin{eqnarray}
H&=&H_0+H_V\nonumber\\
H_0&=&\sum_{i\sigma}^{i=1,2}T_0^dn_{i\sigma}^d+\frac{U}{2}\sum_{i\sigma}^{i=1,2}n_{i\sigma}^dn_{i-\sigma}^d+\sum_k
\epsilon_k^pn_{k\sigma}^p\nonumber\\
H_V&=&V\sum_{i\sigma}^{i=1,2}(d_{i\sigma}^+p_{i\sigma}+h.c.)
\label{H}
\end{eqnarray}
\begin{figure}
\begin{center}
\epsfig{file=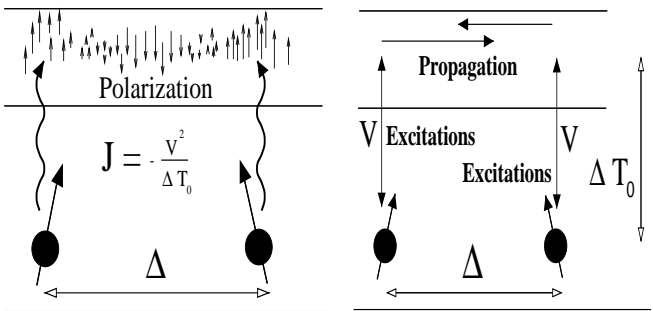, width=1.0\linewidth}
\caption{Schematic picture of conventional RKKY and virtual RKKY
  (superexchange) interaction for large on-site Coulomb repulsion $U\rightarrow\infty$.}
\label{schema}
\end{center}
\end{figure}
Let us note, that the construction operators can stand for holes or for
electrons. If $d^\dagger$ and
$p^\dagger$ create electrons the situation is closest to the usual
interpretation of RKKY-interaction, i.e. that electrons are polarized
by a local spin. For $S=1/2$ the model (\ref{H}) may describe RKKY as
well as superexchange, if the band is partly filled instead of empty. In this case
virtual excitations of band electrons over the Fermi energy contribute
to RKKY, while virtual excitations of the impurity electrons lead to
superexchange (see Fig. \ref{schema}).\\
However in most cases superexchange is constituted by virtual excitations of holes
instead of electrons (e.g. in the "classical" case of MnO). Thus in most
cases the construction operators have to be interpreted as hole creators
and annihilators. $T_0$ and $\epsilon_k$ are now energies for holes and
$U$ is the Coulomb repulsion between holes. The two interpretations of
(\ref{H}) are connected via particle-hole transformation
\begin{equation}
(p,d)^{\dagger}_h\rightarrow (p,d)_e\quad (p,d)_h\rightarrow (p,d)^\dagger_e 
\end{equation}
with the well-known results:
\begin{eqnarray}
T_{0h}^d&=&-(T_{0e}^d+U_e)\nonumber\\
\epsilon_{kh}^p&=&-\epsilon_{ke}^p\nonumber\\
U_h&=&U_e=U\nonumber\\
\mu_h&=&-\mu_e
\label{ph}
\end{eqnarray}
Let us discuss the following situation: The construction operators apply
to holes and thus the hole energies $T_{0h}^d$, $\epsilon_{kh}^p$ and
$\mu_h$ are fixed. Now let us consider the limit $U\rightarrow\infty$. In
this case the one-electron energy of the impurities $T_{0e}^d$ goes to
$-\infty$ while the energy of an doubly occupied impurity orbital stays
finite ($T_{0e}^d+U=-T_{0h}^d$). The same holds for the Bloch energies
$\epsilon_{ke}^p$. For superexchange we want to discuss the parameter
constellation $T_{0h}^d<\mu_h<\epsilon_{kh}^p$ and $U\rightarrow\infty$(see
Fig. \ref{schema}). For electrons this means:
\begin{equation}
T_{0e}^d\ll\epsilon_{ke}^p<\mu_e<(T_{0e}^d+U)
\end{equation}
Hence in the unperturbed ground state the impurity orbitals are filled
with one electron (thus creating local spins). Furthermore the band is
completely filled. Possible excitations are from the band into the
impurity orbital with an excitation energy
\begin{equation}
(T_{0e}^d+U)-\epsilon_{ke}^p=-T_{0h}^d+\epsilon_{kh}^p
\end{equation}
This describes a situation where superexchange is exclusively mediated
by filled (valence) bands.
Such a situation is not only found for the classical magnetic
insulators (MnO), but also in $(II,Mn)VI$ semiconductors \cite{LHE88,LSF88}\\
For $(III,Mn)V$ semiconductors, too, the models of RKKY and superexchange are well
comparable if the construction operators apply to holes, since the
important carriers are holes in these systems.\\
However the simple model (\ref{H}) is quite general and does not only
apply to DMS but to every situation, where virtual excitations of
localized electrons are important.\\
Let us now derive an expression for long range superexchange within this model.
The most instructive way of considering the virtual processes leading to
the spin-spin interaction is perturbation theory since one sums
explicitely over all excited states. Treating the hybridization term as
the perturbation we find that the free ground state is four-fold
degenerated (with respect to the spin configuration). While calculating
the energy corrections it is convenient to characterize the eigen-states
of the free Hamiltonian $H_0$ by their number of
impurity electrons, which is a good quantum number of the free
system. Further the following property of the perturbation $H_V$ should be considered:
\begin{itemize}
\item[] If $H_V$ works on a free ground state it changes the number
  of conduction ($p-$) and impurity ($d-$) electrons (holes) by one (while the total
  number of electrons (holes) is conserved).      
\end{itemize}
Due to this all odd energy corrections $E_a^{(1)}\,E_a^{(3)}\ldots$ vanish\footnote{This is most
  easily seen for the first order energy contribution $\langle
  E_a^{(0)}|H_V| E_a^{(0)}\rangle$. There is the same state at the left
  and the right of $H_V$. Since $H_V$ changes the number of impurity
  electrons by one and free states with a different number of impurity
  electrons are orthogonal this energy correction is zero.}. In second order we find
an energy contribution, which does not affect the degeneracy of the
ground state.
\begin{equation}
E^{(2)}_a=\frac{2}{N}\sum_k \frac{V^2}{T_0^d-\epsilon_k}
\label{2.order}
\end{equation}
The degeneracy is
broken not until fourth order perturbation theory which gives an energy
contribution $E^{(4)}_\alpha$.
Besides a constant term, 
which does not affect the spin orientation, this is given by:
\begin{eqnarray}
E^{(4)}_a=\sum_{bcd}\frac{H_{a}^{b}\cdot
  H_{b}^{c}\cdot H_{c}^{d}\cdot
  H_{d}^{a}}{(E_n^{(0)}-E_m^{(0)})(E_n^{(0)}-E_l^{(0)})(E_n^{(0)}-E_o^{(0)})}\nonumber\\
\label{pert.}
\end{eqnarray}
where $H_x^y=\langle
E_x^{(0)}|H_V|E_y^{(0)}\rangle$.
$|E_{a}^{(0)}\rangle$ is one ground state of the free system with the
ground state energy $E_{n}^{(0)}=2T_0^d$. $|E_{\{b,c,d\}}\rangle$ are
excited eigenstates of the free system with the energies $E_{m}^{(0)}\,
E_{l}^{(0)}$ and $E_{o}^{(0)}$. The sum goes over all excited
states. Due to the special shape of the perturbation potential $H_V$
\begin{figure}
\begin{center}
      \epsfig{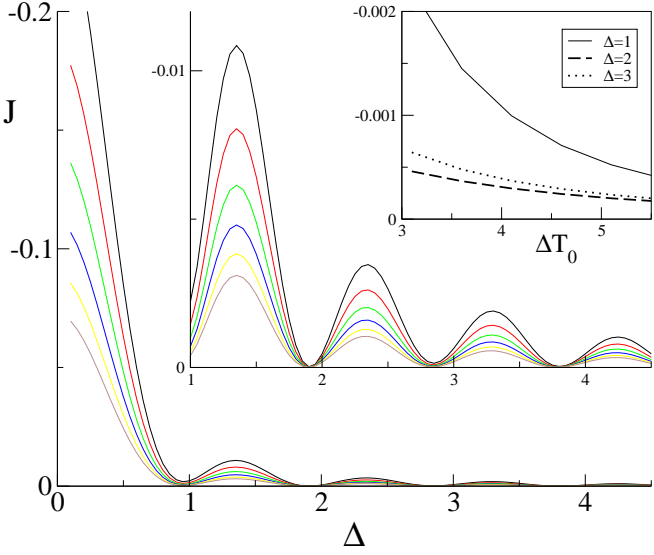}
\caption{Effective spin-spin coupling $J$ as a function of the impurity
  distance $\Delta$ at different energy gaps. The distance is measured in units of the lattice
  constant. The band gaps ($\Delta T_0-\frac{W}{2}$) range from $0.1\;-\;2.6\,eV$ (from top to
  bottom). The impurities are located along $[001]$ of a simple cubic
  tight binding lattice. Other parameters: band-with $W=6\, eV$,
  $V$=0.16$W$. 
Inset: $J$ as a function of the $\Delta T_0$ at (001), (002), (003).} 
\label{sc_3d}
\end{center}
\end{figure}
the latter states must have a certain number of excited electrons
(holes) to get a nonzero energy correction. Since we consider the limit $U\rightarrow\infty$ the excited  
electrons (holes) are in the conduction (valence) band. There is exactly
one electron (hole) in the band in $|E_b^{(0)}\rangle$ and
$|E_d^{(0)}\rangle$ and exactly two electrons (holes) are located within
the band in $|E_{c}^{(0)}\rangle$. Thus $|E_{\{b,c,d\}}\rangle$ can be
expressed in terms of construction operators working on states with two
impurity electrons $|\alpha\sigma_x\beta\sigma_y\rangle$
($\alpha,\beta=1$ or 2; $\sigma=\uparrow$ or $\downarrow$).
\begin{eqnarray}
|E_b^{(0)}\rangle&=&p_{k_1\sigma_1}^\dagger
d_{i\sigma_1}|i\sigma_1\alpha\sigma_x\rangle\nonumber\\
b&=&(k_1\sigma_1\alpha\sigma_x)\nonumber\\
|E_c^{(0)}\rangle&=&p_{k_2\sigma_3}^\dagger p_{k_3\sigma_4}^\dagger
  d_{j\sigma_5}d_{l\sigma_6}|j\sigma_5l\sigma_6\rangle\nonumber\\
c&=&(k_2\sigma_3k_3\sigma_4)\nonumber\\
|E_d^{(0)}\rangle&=&p_{k_4\sigma_7}^\dagger d_{m\sigma_8}|m\sigma_8\beta\sigma_y\rangle\nonumber\\
d&=&(k_4\sigma_7\beta\sigma_8)
\end{eqnarray}   

 After tedious but straightforward calculations one arrives at:
\begin{eqnarray}
E_a^{(4)}=\gamma\sum_{{{k_1\ldots
  k_4\atop o\ldots v}\atop i\ldots m}\atop \sigma_1\ldots \sigma_8}\!\!\!\frac{\langle
E_a^{(0)}|XYZ|E_a^{(0)}\rangle
e^{i\phi}}{(T_0^d-\epsilon_{k_1}^p)(2T_0^d-\epsilon_{k_2}^p-\epsilon_{k_3}^p)(T_0^d-\epsilon_{k_4}^p)}\nonumber
\end{eqnarray}
where $\gamma=\frac{1}{108N^4}$
and
\begin{eqnarray}
X&=&H_Vp_{o\sigma_1}^+(1-n_{i\sigma_2}^d)p_{p\sigma_1}H_Vp_{q\sigma_3}^+p_{s\sigma_4}^+\nonumber\\
Y&=&d_{j\sigma_5}(1-n_{l\sigma_6})d_{j\sigma_5}^+\nonumber\\
Z&=&p_{t\sigma_4}p_{r\sigma_3}H_Vp_{u\sigma_7}^+(1-n_{m\sigma_8}^d)p_{v\sigma_7}H_V
\label{sum}
\end{eqnarray}
\begin{figure}
\begin{center}
\epsfig{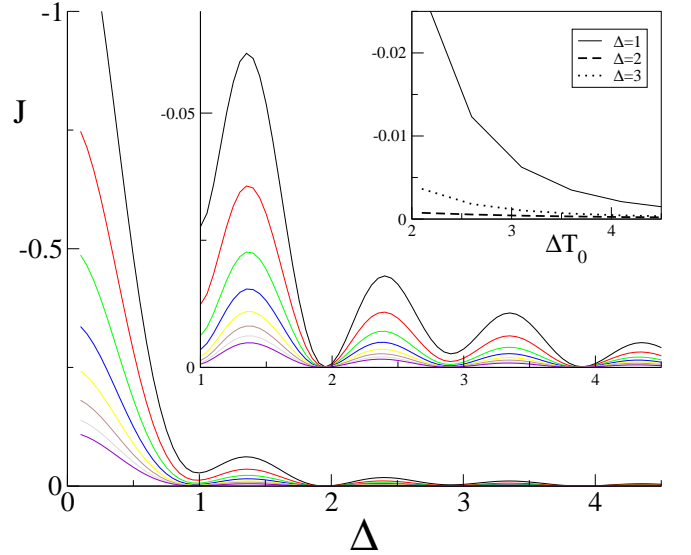}
\caption{As Fig.\ref{sc_3d}, but for a two-dimensional quadratic lattice
  with the band-width $W=4\,eV$, $V=0.25\,W$.}
\label{sc_2d}
\end{center}
\end{figure}
In the sum the subscripts $i\ldots m$ denote impurity sites (1 or 2), while the
subscripts $o\ldots v$ go over all lattice sites. The $k-$ summations
are over the first Brillioune-zone and $\sigma_1\ldots \sigma_8$ are spin-subscripts. The phase factor
$\phi$ reads: $\phi=\bar{k}_1(\bar{R}_o-\bar{R}_p)+\bar{k}_2(\bar{R}_q-\bar{R}_r)+\bar{k}_3(\bar{R}_s-\bar{R}_t)+\bar{k}_4(\bar{R}_u-\bar{R}_v)$. 
Performing the sum and introducing impurity-spin operators as usual
\begin{eqnarray}
S^z_i&=&\frac{1}{2}(n^d_{i\uparrow}-n^d_{i\downarrow})\nonumber\\
S^{(+/-)}_i&=&d_{i(\uparrow/\downarrow)}^+d_{i(\downarrow/\uparrow)}\nonumber
\end{eqnarray}
we finally can write the energy contribution in terms of an effective
Hamiltonian of Heisenberg-form that works on the free ground state $|GS_\alpha^{(0)}\rangle$:
\begin{eqnarray}
E_\alpha^{(4)}&=&\langle GS_\alpha^{(0)}|H_{\rm
  eff}|GS_\alpha^{(0)}\rangle\quad\mbox{with}\nonumber\\
H_{\rm eff}&=&-J(\Delta)\bar{S}_1\cdot \bar{S}_2\quad.
\end{eqnarray}
We find for the exchange integrals $J(\Delta)$ ($\Delta$ is given in
terms of the lattice constant):
\begin{eqnarray}
J=\frac{8V^4}{N^4}\sum_{k_1\ldots
  k_4}\frac{F(\Delta)}{(T_0^d-\epsilon_{k_1}^p)(2T_0^d-\epsilon_{k_2}^p-\epsilon_{k_3}^p)(T_0^d-\epsilon_{k_4}^p)}\nonumber\\
{}\nonumber\\
F(\Delta)=2\cos((k_2-k_3)\Delta)+4\cos((k_1-k_2)\Delta)\qquad\qquad\nonumber\\
+\cos((k_1-k_4)\Delta)+\cos((k_1+k_4-k_2-k_3)\Delta)\qquad
\label{J}
\end{eqnarray}
This effective spin coupling is of the anticipated order
$\frac{V^4}{\Delta T_0^3}$ for small distances $\Delta$. Due to
the four-fold sum in (\ref{J}) we can not give an analytic expression
for the asymptotic behaviour of $J(\Delta)$. However, since excitations
over the band gap are necessary we expect an exponential decay. 
In contrast to the "classical superexchange" where the particles fluctuate between the
impurities and a single degenerated intermediate state, now the electrons
may hop into different Bloch-states and still cause an effective
interaction.\\
For the zero-bandwidth limit, i.e. $\epsilon_k^p=T_0^p$ for
all $k$, the $k-$sum in (\ref{J}) goes only over $F(\Delta)$.
Since each cosine-function adds now to zero, the interaction vanishes in
this limit. This is the correct result, because the sites are
completely decoupled in the zero-bandwidth limit.\\
The numerical evaluation of Eq. (\ref{J})gives always an
anti-ferromagnetic interaction that declines with the distance and shows
certain oscillations (see Fig. \ref{sc_3d}).
 The interaction
gets even more important for systems with reduced dimensionality. For a two
dimensional lattice the magnitude of the interaction increases
approximately by a factor of five.
 This is seen in Fig. \ref{sc_2d}, where we used the same parameters for the nearest
neighbor hoppings and the gaps between the band and the
impurity-level as in Fig. \ref{sc_3d}.\\
Eq. (\ref{J}) gives an exact result in perturbation
theory for a well-defined limit. Other treatments of long range superexchange
that may involve more complicated models but also some additional
approximations can be compared in the limit $U_h\rightarrow\infty$ with
Eq. (\ref{J}).\\ Let us demonstrate this with an example in literature that
treats $(II,Mn)VI$ semiconductors. In Ref. \cite{LHE88} Larson
et.al. investigate electron and hole mediated superexchange and a special kind of
Bloembergen-Rowland interaction (negative local $J$). They apply a
multi-band model with a realistic electronic structure, a
local Coulomb repulsion $U$ between $Mn-3d$ electrons and a
hybridization between Mn ions and the host material. As explained in
their paper, the five degenerated $Mn-d$ orbitals can be modelled by a
single orbital plus a factor that depends on the Mn ground state
only. Thus only a single orbital is considered at each Mn site. The
authors found that the superexchange caused by virtual excitations of two holes
dominate. After applying the limits $U\rightarrow\infty$, single nondegenerate valence band and local
hybridization ($V(k)=V$) to their result, we want to compare them with Eq. (\ref{J}). The result of Larson
et.al. (Eq.(4.4) of Ref. \cite{LHE88}) is written with electronic
parameters. To compare it with our result we have to perform a particle
hole transformation (\ref{ph}) and apply the just mentioned limits and
simplifications. Then the Mn-Mn exchange of Ref.\cite{LHE88} reduces to:
\begin{equation} 
J_{hh}^{dd}(\Delta)=2\sum_{kk^\prime}\frac{V^4\cos(k-k^\prime)\Delta}{(T_0^d-\epsilon_k)^2(T_0^d-\epsilon_{k^\prime})}
\end{equation}
This is quite close to the exact result in fourth order perturbation
theory (Eq.(\ref{J}). The remaining discrepancies seem to be a fair
price for the complexity of the model investigated in
Ref. \cite{LHE88}.\\
Finally let us discuss qualitatively the influence of free carriers on
superexchange and RKKY. If free carriers are doped into the band the virtual
excitations of this carriers over the Fermi energy lead to RKKY
interaction. Since the energy gap is much smaller for these carriers the
RKKY contribution should dominate in the sum (\ref{pert.}). Furthermore,
since the band is now partly occupied there are less virtual
intermediate states for superexchange. This gives a vivid explanation for the fact
that superexchange is suppressed by free carriers as e.g. worked out by Qimiao Si
et.al. for CuO (Fig.1 of Ref. \cite{SLL92})\\
In conclusion we have derived a simple expression for long range superexchange in
a well defined toy model. This expression is useful for qualitative
discussions and constitutes a limit, which can be used to evaluate
approximations in more complex models. We have given an example of one such
comparison for the case of $(II,Mn)VI$ semiconductors, where the long
range component of superexchange is very interesting. However, as in the case of
RKKY, the physical picture developed here is quite general and is
applicable to all problems where virtual excitations of two electrons or
of two holes lead to an effective spin-spin coupling between these
electrons or holes.\\
This work was supported by the Deutsche Forschungsgemeinschaft within
the Sonderforschungsbereich 290.

\end{document}